\documentclass[acmtog,nonacm]{acmart}

\AtBeginDocument{%
  \providecommand\BibTeX{{%
    \normalfont B\kern-0.5em{\scshape i\kern-0.25em b}\kern-0.8em\TeX}}}

\setcopyright{none}

\acmJournal{TOG}
\acmVolume{37}
\acmNumber{4}
\acmArticle{111}
\acmMonth{8}

\begin{document}

\title{Optimal Dual Schemes for Adaptive Grid Based Hexmeshing}

\author{Marco Livesu}
\orcid{0000-0002-4688-7060}
\email{marco.livesu@gmail.com}
\affiliation{%
  \institution{CNR IMATI, Italy}
}

\author{Luca Pitzalis}
\affiliation{%
  \institution{University of Cagliari and CRS4, Italy}
}

\author{Gianmarco Cherchi}
\orcid{0000-0003-2029-1119}
\affiliation{%
  \institution{University of Cagliari, Italy}
}

\renewcommand{\shortauthors}{Livesu, M et al.}

\begin{teaserfigure}
\includegraphics[width=\linewidth]{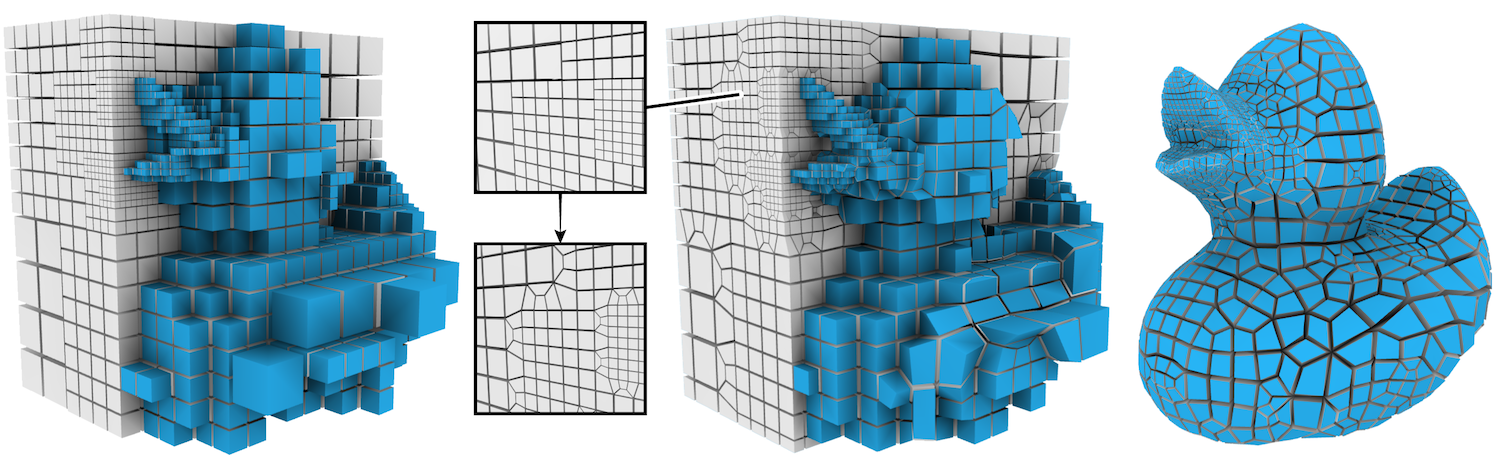}
\caption{We propose a novel set of dual schemes to turn an adaptively refined grid into a conforming hexahedral mesh. Compared with prior methods, our schemes allow to process a broader class of input grids, and produce hexahedral meshes with a simpler singular structure.}
\label{fig:teaser}
\end{teaserfigure}

\begin{abstract}
  Hexahedral meshes are an ubiquitous domain for the numerical resolution of partial differential equations. Computing a pure hexahedral mesh from an adaptively refined grid is a prominent approach to automatic hexmeshing, and requires the ability to restore the all hex property around the hanging nodes that arise at the interface between cells having different size. 
  The most advanced tools to accomplish this task are based on mesh dualization. These
  approaches use topological schemes to regularize the valence of inner vertices and edges, such that dualizing the grid yields a pure hexahedral mesh.
  In this paper we study in detail the dual approach, and propose four main contributions to it: (i) we enumerate all the possible transitions that dual methods must be able to handle, showing that prior schemes do not natively cover all of them; 
  (ii) we show that schemes are internally asymmetric, therefore not only their implementation is ambiguous, but different implementation choices lead to hexahedral meshes with different singular structure; (iii) we explore the combinatorial space of dual schemes, selecting the minimum set that covers all the possible configurations and 
  also yields the simplest singular structure 
  in the output hexmesh; (iv) we enlarge the class of adaptive grids that can be transformed into pure hexahedral meshes, relaxing one of the tight requirements imposed by previous approaches, and ultimately permitting to obtain much coarser meshes for same geometric accuracy.
  Last but not least, 
  for the first time we make grid-based hexmeshing truly reproducible, releasing our code and also revealing a conspicuous amount of technical details that were always overlooked in previous literature, creating an entry barrier that was hard to overcome for practitioners in the field.
  \end{abstract}



\maketitle

\section{Introduction}

\label{sec:intro}









%



Volumetric discretizations made of hexahedral cells are ubiquitous in applied sciences, where they are used as computational domains for the numerical resolution of partial differential equations~\cite{schneider2019poly,wang2021comparison,wang2004back}.
Converting an adaptively refined grid into a pure hexahedral mesh is one of the prominent approaches to hexmeshing, and due to its unbeaten scalability and robustness is the only fully automatic method that successfully transitioned from academic research to an industrial software~\cite{meshgems}.

Grid-based methods operate on carefully constructed adaptive lattices. Hanging nodes arising at the interface between cells with different size are substituted by dedicated topological schemes that locally restore the all hex connectivity and provide the necessary mesh grading. For ease of implementation and reproduction, the number of these schemes must be low. At the same time, the scheme set must be rich enough to exhaustively address all the possible configurations, such that convergence to a pure hexahedral mesh is always guaranteed.\\

\begin{figure*}
	\centering
	\includegraphics[width=\linewidth]{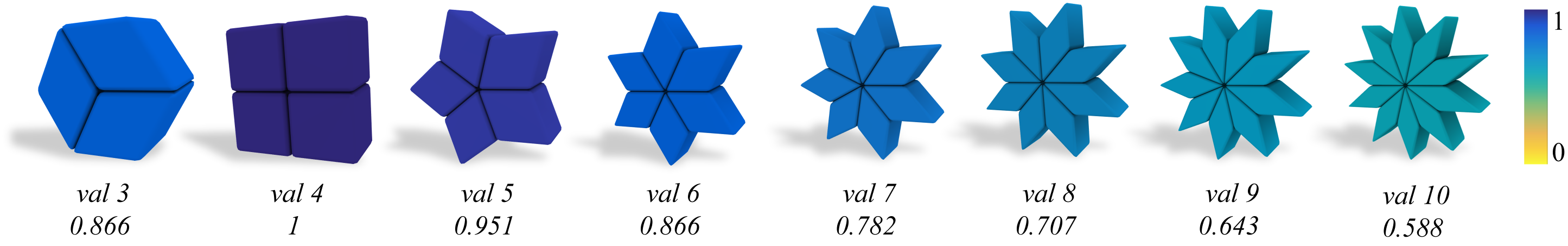}
	\caption{The number of hexahedra incident to an edge directly impacts the maximum per element quality locally achievable. For each configuration, we show edge valence and Scaled Jacobian, optimized with~\cite{livesu2015practical}. Our schemes always generate edges with valence between 3 and 6. On the dataset released with~\cite{gao2019feature}, both~\cite{marechal2009advances} and~\cite{gao2019feature} introduce singular edges with valence up to 9 and 10, respectively.}
	\label{fig:valence}
\end{figure*}

Early approaches unsuccessfully tried to directly incorporate hanging nodes in the hexmesh, but this operation is not always possible (Section~\ref{sec:related}). Marechal~\shortcite{marechal2009advances} was the first to observe 
that if all grid vertices have valence six and all edges have four incident cells, then the dual of the grid mesh yields only hexahedral elements. 
The schemes he proposed operate on small groups of nearby hanging nodes, regularizing their valence. In 2D a single scheme is sufficient to handle all the possible cases.
 For 3D, he proposed three alternative schemes for flat, convex, and concave transitions.\\

In this paper, we dive into the details of the dual approach. This research was initially motivated by the mere necessity to replicate the results showcased in~\cite{marechal2009advances,marechal2016all}. However, our analysis soon revealed that the ideas of Marechal were somehow incomplete and ambiguous, and that novel contributions were necessary to complete the picture and make grid-based hexmeshing exhaustive and fully reproducible. Specifically, this article offers four main contributions:

\begin{figure*}[t]
	\centering
	\includegraphics[width=.9\linewidth]{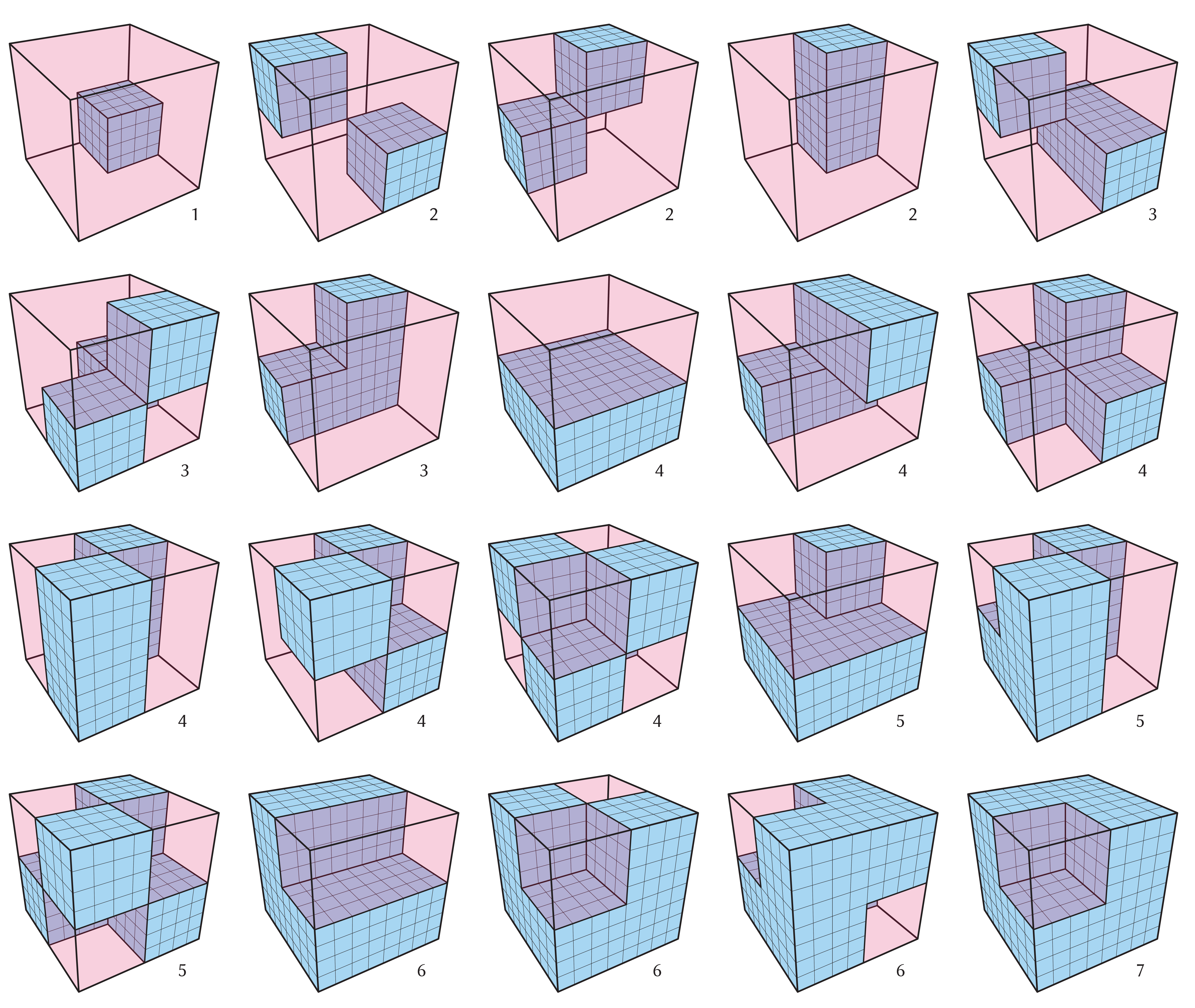}
	\caption{Exhaustive set of all possible transitions that may arise in an adaptive grid with two alternative levels of refinement. Values next to each cube denote the number of octants filled with a regular $4 \! \times 4 \!\times 4$ sub-grid (blue). The red-shaded empty octants can be imagined filled with a coarser $2 \! \times 2 \!\times 2$ sub-grid, which is not rendered to make the figure easier to parse. These cases can be seen as a volumetric interpretation of the lookup table shown in~\cite{nielson2004dual}.}
	\label{fig:20cases}
\end{figure*}

\begin{itemize}

	\item  we show that the original schemes proposed in~\cite{marechal2009advances} and subsequent works~\cite{gao2019feature,hu2013adaptive} are not exhaustive, in the sense that they only show the basic transitions, without explicitly deriving the full scheme set that is necessary to handle all the possible configurations. Specifically, considering the class of adaptive grids that can be processed with these methods, there exist exactly 20 alternative transitions (Figure~\ref{fig:20cases}). Handling all of them requires a non trivial blending of the known elementary schemes, which was never discussed in previous literature;\\

	\item we show that prior schemes are ambiguous, because  
	transitions are internally made of chains of prismatic elements that intersect orthogonally, passing one on top of the other. It follows that
	schemes are internally asymmetric, and there are always two possible ways to handle an intersection. In flat regions two chains intersect once, hence there are two alternative ways to implement them. Concave edges have three chains that intersect twice, hence there are $2^2$ configurations. Concave corners involve three chains that intersect three times, hence there are $2^3$ configurations. Interestingly, these choices are often critical, as they impact the singular structure of the output mesh;\\

	\item we explore the combinatorial space of dual schemes, proving that there are multiple ways to bend a chain of prisms around a concavity, and that each method produces a dual hexmesh with different singular structure. We also show that we can avoid high valence edges by wisely selecting the best combinations, producing meshes having only singular edges with valence 3, 5, and 6. Existing tools based on~\cite{marechal2009advances,gao2019feature} insert edges with higher valence, negatively affecting mesh quality (Figure~\ref{fig:valence});\\

	\item we extend the class of adaptive grids that can be transformed into pure hexahedral meshes. Prior schemes require the input grid to be (strongly) balanced, meaning that the difference in the amount of refinement associated with any pair of cells sharing an edge, face or vertex must be lower than 2. We introduce a few additional schemes that permit input grids to obey to a weaker definition of balancing, where only face-adjacent cells must have compatible refinement. This extension allows to greatly reduce the number of mesh elements, without affecting geometric accuracy (Section~\ref{sec:bal_comparisons}).

\end{itemize}

Summarizing, this study offers a comprehensive overview -- and hopefully a better understanding -- of the state of the art in grid-based hexmeshing, also proposing novel ideas and advancing the field. Our first practical contribution amounts to a set of 8 schemes that exhaustively address all the 20 transitions for strongly balanced grids. We do not claim that we are the firsts to secure a conversion into a pure hexahedral mesh (both the commercial implementation of~\cite{marechal2009advances} and the software released with~\cite{gao2019feature} can do this), but our schemes are the firsts that make this conversion easy, also producing hexahedral meshes with a superior singular structure (Section~\ref{sec:comparisons}).
Our second practical contribution amounts to 5 additional schemes that permit to handle the transitions that arise in weakly balanced grids. To the best of our knowledge, this extension was never attempted before.
To grant maximal diffusion, we publicly share the complete set of schemes and the code necessary to install them. All these contributions have been incorporated into the MIT licensed library CinoLib~\cite{Liv19}.




\section{Related works}
\label{sec:related}

Grid-based hexahedral meshing was pioneered by Schneiders, who firstly proposed to use regular voxel grids~\cite{schneiders1996grid}, and soon later introduced adaptively refined grids, obtained through the use of octrees~\cite{schneiders1996octree}. Octrees had already been used for adaptive mesh generation, but they were unsuitable to hexmeshing because there were no topological schemes to suppress hanging nodes, and there were no bounds on the topological complexity of each cell~\cite{shephard1991automatic}.

If adjacent elements in an adaptive grid differ by at most one level of refinement, there exist exactly $2^8$ alternative configurations that must be handled. Discarding trivial configurations (i.e., fully refined and unrefined cells) and accounting for symmetries, the number goes down to 20 unique cases~\cite{weiler1996automatic}. A hexahedralization for four of these configurations was published multiple times~\cite{tack1994two,schneiders1996octree,schneiders2000octree,schneiders2000algorithms,schneiders1997algorithm}. Authors also showed that concave configurations cannot be hexmeshed, because they contain an odd number of quadrilateral faces, a condition for which it is proved that the hexmeshing problem does not admit a valid solution~\cite{mitchell1996characterization}. Despite non-exhaustive, the known schemes are of practical relevance, and a variety of algorithms used them to generated adaptive hexahedral meshes. Whenever unsupported configurations arise, grid refinement can be iteratively applied to reconduct the problem to the set of known schemes so that a mesh can be created~\cite{zhang2006adaptive}. This allows to potentially mesh any grid, but at the same time is undesirable, because a conspicuous amount of extra refinement may be necessary to converge to a valid solution. Slightly different refinement schemes were proposed in~\cite{ito2009octree}, but also in this case concave configurations cannot be handled. Conversely, \cite{LMPS16} proposed schemes to realize fine-to-coarse transitions along a tubular object, but this approach does not extend to a broader class of shapes.

Considering the impossibility to derive a complete set of hexmeshing schemes, recent literature approached the problem from a different angle. Marechal~\shortcite{marechal2009advances,marechal2016all} pioneered the dual approach, which is based on the intuition that cutting grid cells to regularize vertex valences defines a polyhedral mesh that yields a pure hexahedral mesh once dualized. 
His paper sets the basic ideas but does not precisely describe the schemes, at the point that we faced major difficulties in reproducing them (Section~\ref{sec:comparisons}). Based on the same principles, we independently derived an optimal set of schemes, for which we demonstrate both exhaustiveness and optimality, in the sense that they produce meshes with simpler singular structure, containing only edges with valence in between 3 and 6.

In recent years a few variations of the dual approach have been proposed. Hu and colleagues~\shortcite{hu2013adaptive} propose to position transition schemes at the inner sides of refined areas, reducing their volumetric extent. The idea is to essentially shift the same schemes one level inwards in the grid, without changing them. Gao and colleagues~\shortcite{gao2019feature} proposed to dualize the grid first, and then substitute clusters of non-hexahedral elements with templated all-hex schemes which reproduce patterns very similar to the ones designed by Marechal.
In Section~\ref{sec:comparisons} we compare our schemes with~\cite{marechal2009advances,marechal2016all,gao2019feature}, showing that for strongly balanced grids our transitions produce a superior singular structure in the output hexmesh. None of these prior methods supports weakly balanced grids.

\paragraph{Other techniques.} 
Hexahedral meshing is a vast topic, and a variety of alternative techniques have been proposed in literature, such as polycubes~\cite{LVSGS13,fang2016all}, advancing front methods~\cite{kremer2014advanced}, sweeping methods~\cite{gao2015structured}, and methods that align to some guiding field~\cite{LPPSC20,liu2018singularity,Corman:2019:SMF}. As of today, none of these approaches can be compared to grid based techniques in terms of scalability and robustness. Despite aiming to produce the same type of meshes, there are no methodological overlaps with these techniques, which are unrelated to our work.
\section{Dual meshing: constraints and desiderata}
\label{sec:dualization}
The dual idea is a broad topological concept, with applications in many scientific fields. In mesh generation dualization has been widely used, e.g. to transform a Voronoi diagram into a simplicial mesh~\cite{levy2010p}, or to generate quadrilateral~\cite{nielson2004dual,campen2012dual} and hexahedral~\cite{tautgesa2003topology,marechal2009advances,hu2013adaptive,gao2019feature} meshes.

Considering a (primal) cellular complex composed of $V$ vertices, $E$ edges, $F$ faces and $C$ cells, its dual mesh is a cellular complex having:
\begin{itemize}
	\item one vertex for each primal cell $c \in C$
	\item one edge for each primal face $f \in F$
	\item one face for each primal edge $e \in E$
	\item one cell for each primal vertex $v \in V$
\end{itemize}
In particular, the valence of each dual vertex corresponds to the number of faces of its associated primal cell. The valence of each dual edge corresponds to the number of sides of its primal face. The number of sides of each dual face corresponds to the valence of its associated primal edge. The number of faces of each dual cell corresponds to the valence of its associated primal vertex.

From a topological perspective, a hexahedron is a solid with 8 vertices, 12 edges, and 6 quadrilateral faces. Considering the definition above, one can always generate a pure hexahedral mesh via dualization if and only if:
\begin{itemize}
	\item each primal vertex has valence 6, because its associated dual cell has 6 faces
	\item each primal edge has valence 4, because its associated dual face will be a quad
\end{itemize}

In addition to these strict topological requirements, it is practically relevant to ensure that the so generated hexmesh has a good singular structure, meaning that it locally resembles a regular grid almost everywhere. To this end, it is desirable that the majority of inner dual vertices have valence 6, and that the majority of dual edges have valence 4. Thinking about these properties in terms of their relation with the primal mesh, it turns out that a good adaptive grid should have as many cells as possible composed of 6 faces, and as many 4-sided faces as possible. In particular, it is important to avoid primal faces with many sides, because they produce high valence singular edges in the dual hexmesh, which negatively impact per element quality (Figure~\ref{fig:valence}). The schemes proposed in this paper are designed to fully address topological constraints, and also to optimize the fulfillment of practical desiderata.
\begin{figure}[htb]
	\centering
	\includegraphics[width=\columnwidth]{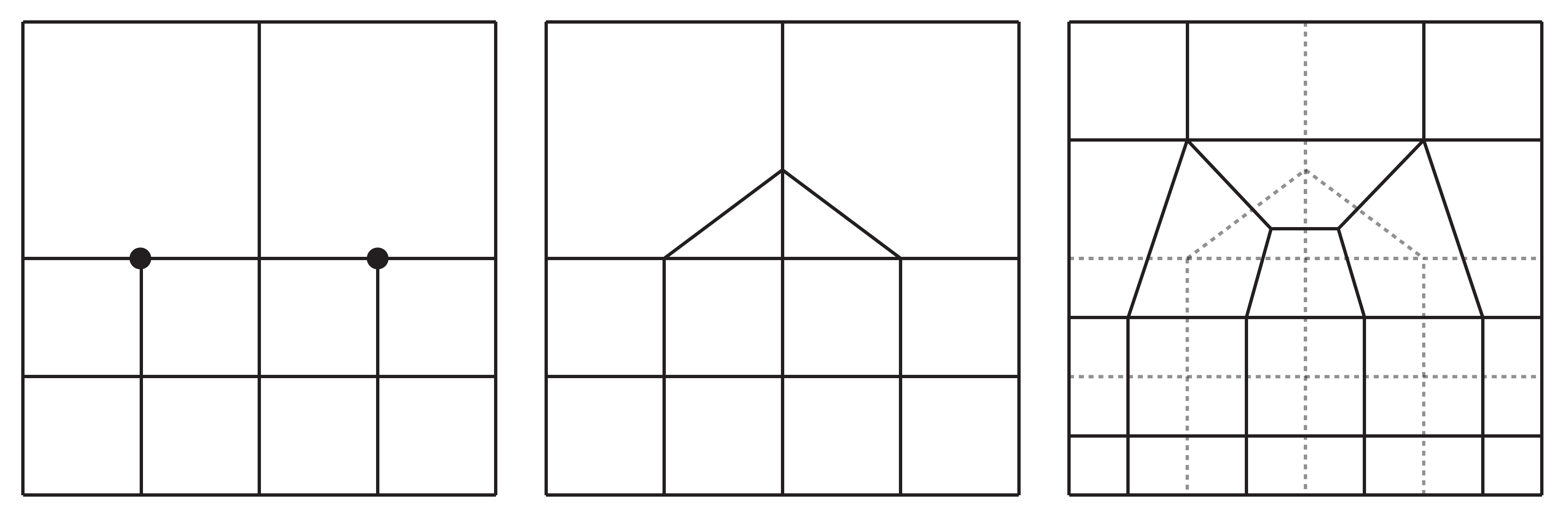}
	\caption{A 2D example of the dual approach proposed in~\cite{marechal2009advances}. Left: an adaptively refined grid has two hanging vertices (black dots) at the interface between elements of different sizes. Middle: connecting them through the vertical edge in between ensures that all internal nodes have valence four. Right: dualizing the grid yields a pure quadrilateral mesh.}
	\label{fig:marechal2d}
\end{figure}

\begin{figure*}[t]
	\centering
	\includegraphics[width=\linewidth]{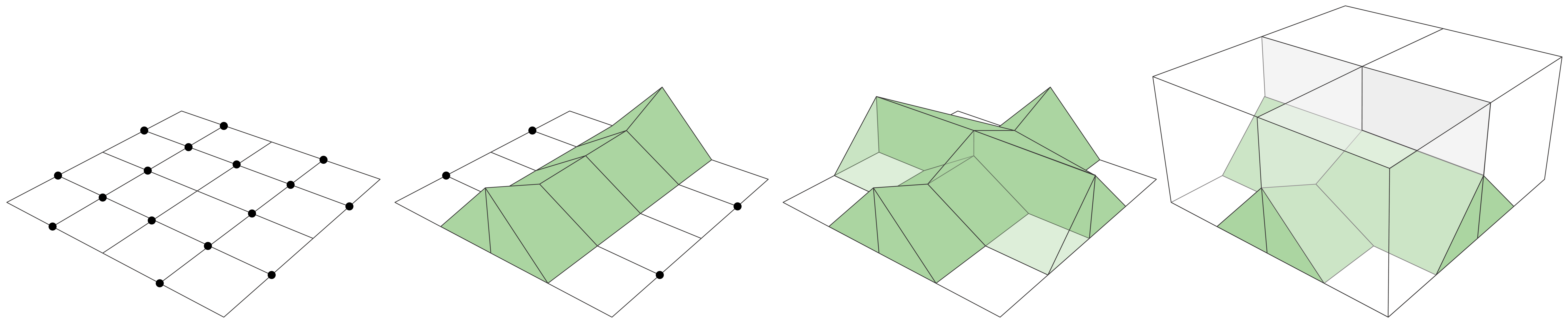}
	\caption{Basic scheme to transition from a flat $4 \!\times\!4$ to a $2 \!\times\!2$ grid. From left to right: there are 14 hanging nodes with valence 5 (black dots). The first chain of prismatic elements with triangular cross section suppresses all hanging nodes but four. The second chain intersects the first one orthogonally, and secures valence 6 for all vertices. The four upper cells reproduce the $2 \!\times\!2$ structure, completing the transition. Some of the faces are transparent to better inspect the interior topology.}	
	\label{fig:flat}
\end{figure*}

\section{Basic Transitions}
\label{sec:basic_scheme}
In this section we start from the basic scheme originally proposed in~\cite{marechal2009advances} and show how it can be adapted to convex, concave, and corner configurations. As it will become clear in the remainder of the section, there are multiple ways to perform this task. We will exhaustively show all the possible versions and select the ones that are optimal with respect to the desiderata expressed in Section~\ref{sec:dualization}.

The core 
idea is identical to the 2D case depicted in Figure~\ref{fig:marechal2d}, where the two hanging nodes are suppressed by forming two triangles connected through the vertical edge in between them. However, the 3D realization is more convoluted, because any non conforming transition between a $4 \!\times\! 4$ and a $2 \!\times\!2$ grid generates 14 hanging vertices with valence 5. 
Following the analogy with the 2D case, we can imagine to extrude the triangles that suppress the hanging nodes, which become chains of prismatic elements with triangular cross-section. Each transition requires two such chains, that intersect orthogonally at the middle of the grid. If the chains were identical, their intersection would define a valence 8 vertex, violating the constraints expressed in Section~\ref{sec:dualization} and thus failing to produce a pure hexahedral dual mesh. To keep vertex valences under control, the trick is to make sure that the two chains 
do not intersect at the same height, but rather one passes below the other, splitting the valence 8 vertex into two valence 6 vertices. Consequently, any time there is an intersection, one must choose which of the two chains passes below the other, leading to ambiguity. Figure~\ref{fig:flat} shows how the topology of the two chains must be arranged to secure the correct vertex and edge valences. One could alternatively choose to have the upper chain passing below the lower one. This choice is completely harmless if the intersection occurs at a flat region, because the global amount and type of primal mesh elements does not change.\\

Given this basic scheme, the whole idea behind dual hexmeshing is to suppress all hanging nodes by designing a network of prismatic chains that wind around clusters of grid elements having the same amount of refinement. Since each cluster is a regular sub-grid, its outer surface is also regular, therefore chains always intersect pairwise in an orthogonal manner. In practice, this means that all we need is to be able to adapt the scheme in Figure~\ref{fig:flat} to allow these chains to turn at the convexities and concavities of each refined cluster.

%


\subsection{Convex transitions}
\label{sec:convex_edge}
\begin{figure}[t]
	\centering
	\includegraphics[width=\linewidth]{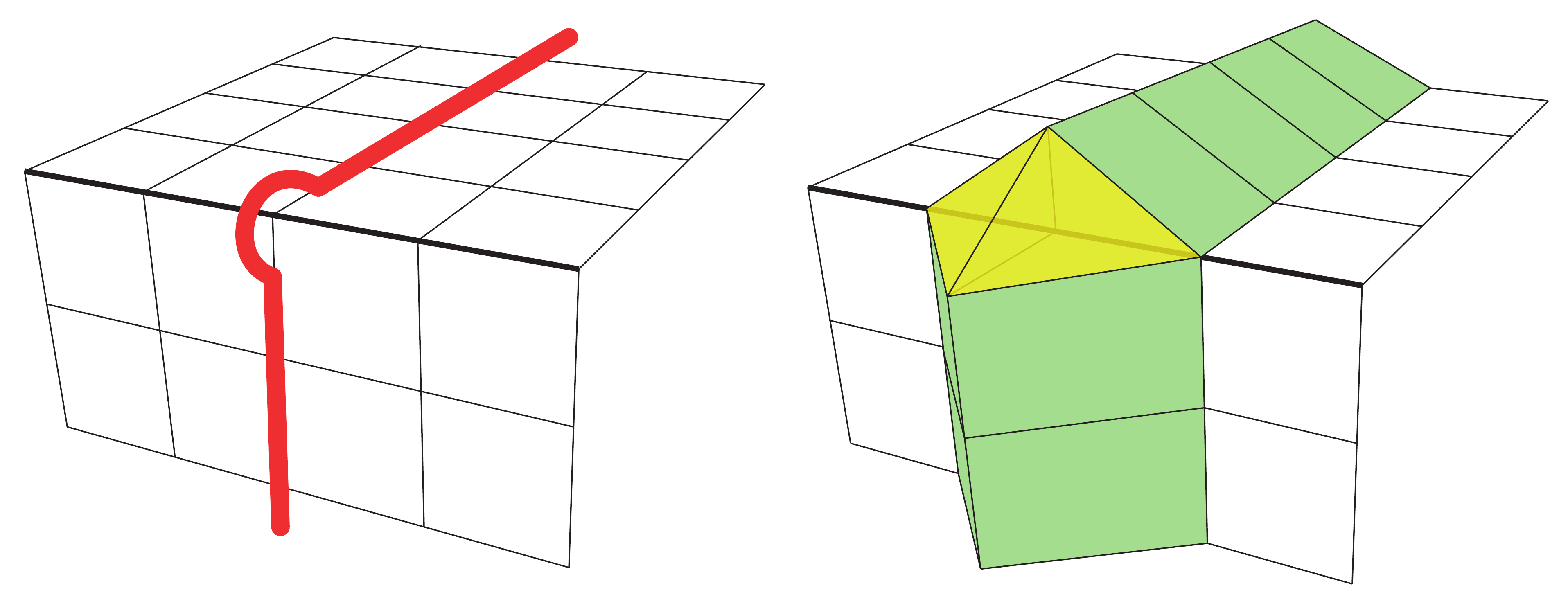}
	\caption{When a chain of prismatic elements turns $90^\circ$ to traverse a convex edge, two tetrahedral elements (yellow) are necessary to adjust mesh topology and provide the necessary bending.}
	\label{fig:convex}
\end{figure}

Two chains of prisms that meet at the convex edge of a refined area can be welded together by using two tetrahedral elements that form a bridge between the cross sections of the incoming chains (Figure~\ref{fig:convex}). Differently from flat and concave transitions, this scheme is not ambiguous because no intersections between orthogonal chains are involved.

\begin{figure*}[t]
	\centering
	\includegraphics[width=\linewidth]{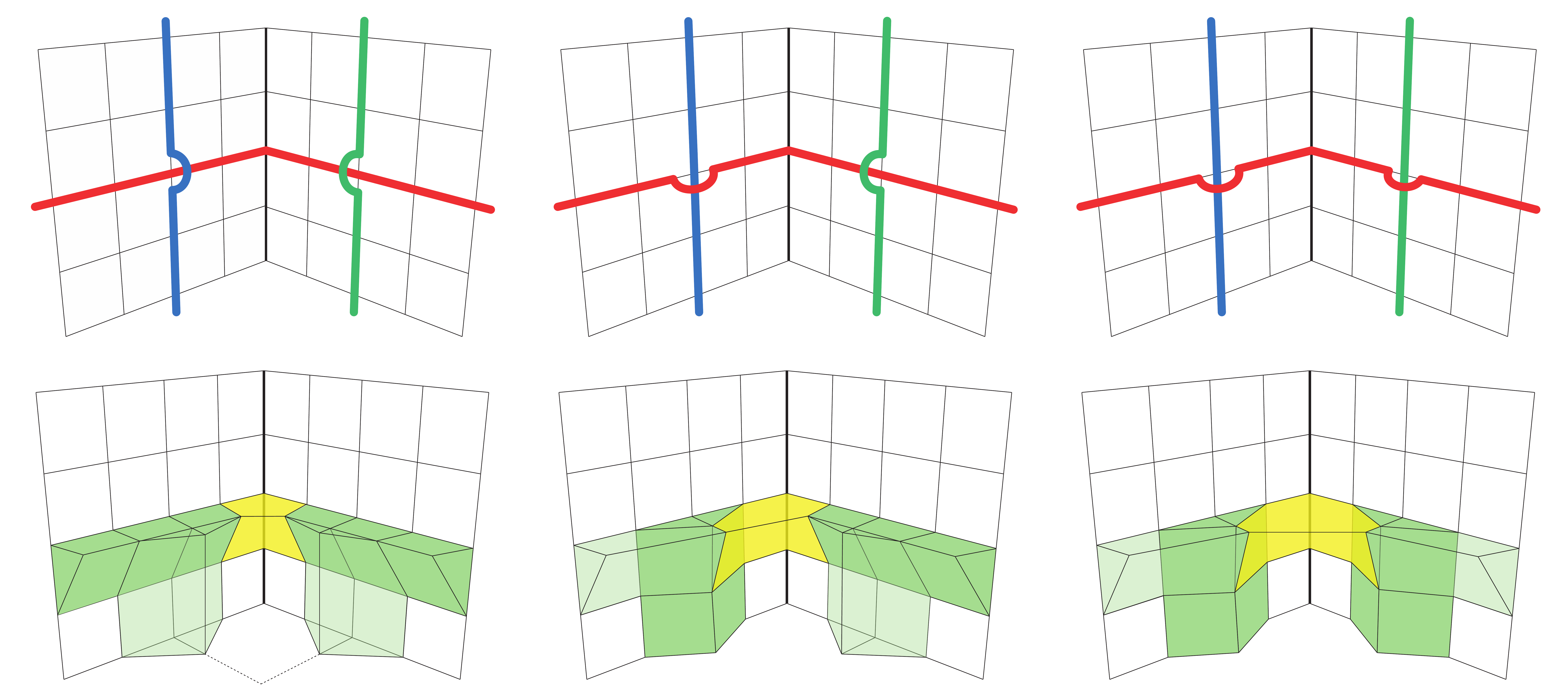}
	\caption{There are three different ways to bend a chain of prismatic elements along a concave edge. Left: if the chain passes below both intersections aside the concavity, the bending can be realized with three pentagonal faces. Middle: if the chain passes below one intersection and above the other, the bending necessitates hexagonal faces. Right: if the chain passes above both intersections, heptagonal faces are needed. Transition elements are highlighetd in yellow. Note that in all cases two hexagonal faces are needed to handle the bottom and top corner faces (see the bottom left dashed lines). The leftmost solution is the optimal one, because it introduces the least amount of high valence irregular edges in the dual hexmesh.}
	\label{fig:edge_configurations}
\end{figure*}

\subsection{Concave transitions}
\label{sec:concave_edge}
Transitions across concave edges are more complex, because four different chains of prismatic elements are involved. Two of them are parallel to the concave edge, and are positioned aside from it. The other two are orthogonal to the concave edge, and are the ones that need to be merged into a single chain that turns at the concavity. These four chains intersect pairwise at the left and right of the concave edge. Depending on how these intersections are realized, the transition changes. There are two different ways two handle each intersection (one chain goes below, one above), therefore there exist $2^2$ alternative solutions. Ignoring symmetries, the amount of unique schemes reduces to three. Specifically, if the two chains that merge at the concave edge pass both below their respective intersections, their blending can be realized using three pentagonal faces (Figure~\ref{fig:edge_configurations}, left). If one of the two chains passes above its intersection, then three hexagonal faces are needed (Figure~\ref{fig:edge_configurations}, middle). Finally, if both chains pass above their intersections, three heptagonal faces are needed (Figure~\ref{fig:edge_configurations}, right).


Recalling that primal faces become edges in the dual mesh, and that the valence of such edges correspond with the number of sides of their primal face, we can conclude that --  depending on the scheme of choice -- concavities may introduce irregular edges with valence 5, 6 or 7. In order to optimize the criteria expressed in Section~\ref{sec:dualization} we always adopt the transition that produces valence 5 edges, obtaining the simplest singular structure in the output mesh. Note that regardless of the configuration of choice, the top and bottom lids of a concave edge are essentially two quads with two corners cut (to account for the incoming chains). This means that the full scheme will still produce two valence 6 edges in the dual mesh (see the dashed lines at the bottom left of Figure~\ref{fig:edge_configurations}). Nevertheless, our choice minimizes the extent of high valence singularities in the output hexmesh, completely avoiding valence 7 edges and reducing the amount of valence 6 singular edges to only two.



\subsection{Transitions around corners}
\label{sec:corners}
\begin{figure}[t]
\centering
\includegraphics[width=\columnwidth]{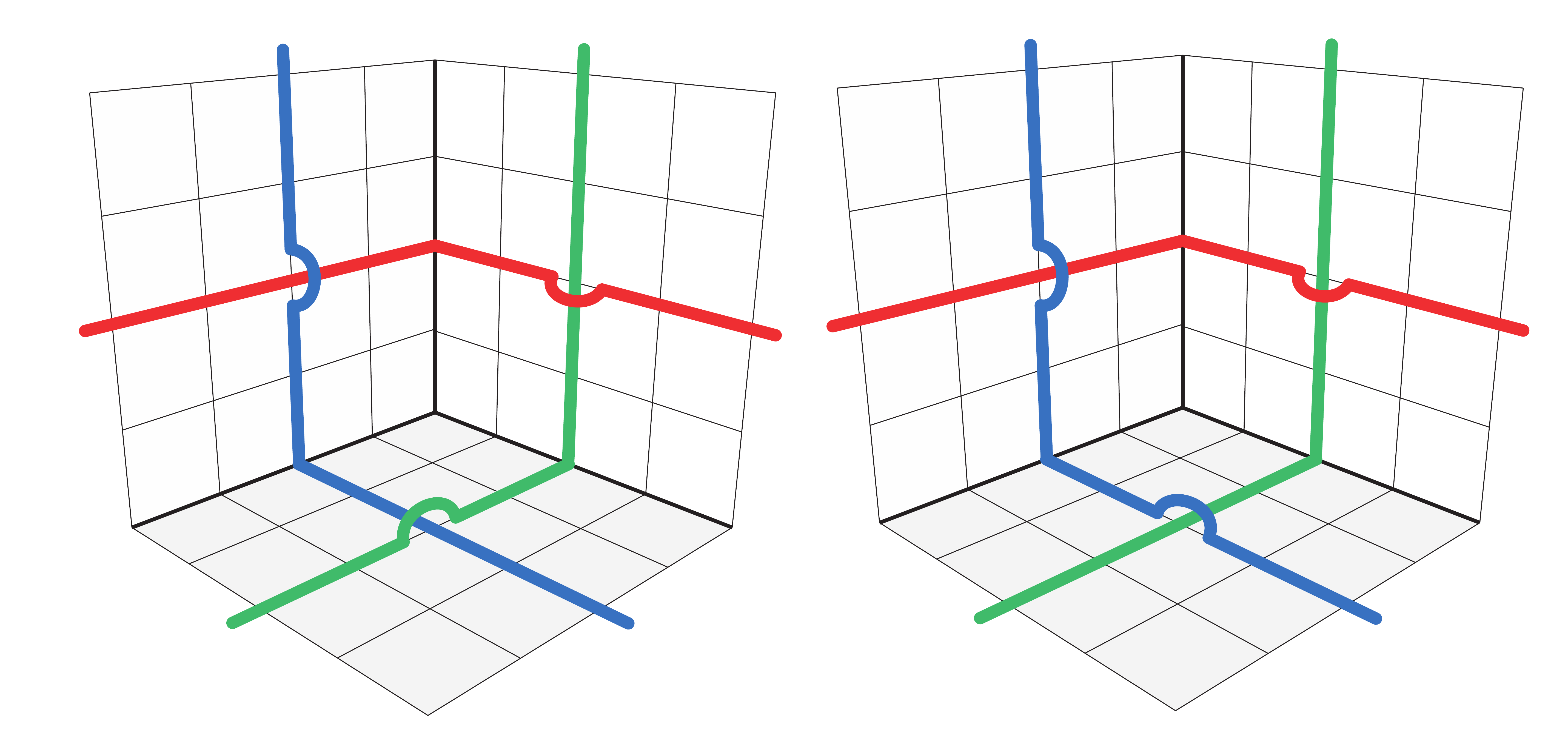}
\caption{Corners of a refined area always have three chains of prisms winding around them. Since each intersection chain can be arranged in two different ways, there are $2^3$ alternative configurations, which remove symmetries to the two shown in this figure. The left configuration is fully symmetric, as the chain traversing each concave edge passes above one intersection and below the other one. The right configuration exposes all the three possible cases (one chain fully below, one chain fully above, and one chain both below and above). The leftmost configuration is better because it only introduces valence 6 edges in the hexmesh, whereas the one at the right also introduces valence 7 singular edges.}
\label{fig:corner_configurations}
\end{figure}

Prismatic chains never traverse the corners of a cluster of refined elements directly, but each corner has three chains that wind around it and mutually intersect each other three times. If the corner is convex, these intersections are handled with the flat scheme in Figure~\ref{fig:flat}, and always produce a mesh with equivalent singular structure. 
Conversely, concavities require to use a blending of the schemes for concave edges shown in Figure~\ref{fig:edge_configurations}. Differently from a single concavity, which can always be handled with the simplest among the three possible options, concave corners are the meeting point of three mutually orthogonal concave edges. The interplay between the chains winding around the corner is such that it becomes impossible to make sure that each chain passes below all intersections it is involved in.
More precisely, three mutual intersections and two alternative ways to handle each one of them define a combinatorial space of $2^3$ alternative solutions. Removing the symmetries, there exist only two ways to handle a concave corner: in one case, each concave turn involves a chain that passes above one intersection and below the next one (Figure~\ref{fig:corner_configurations}, left). In the other case, all the three transitions shown in Figure~\ref{fig:edge_configurations} arise. Also in this case, our preference goes to the left configuration because it fully avoids the generation of singular edges with valence 7 in the output hexmesh and minimizes the amount of valence 6 edges.

\section{Schemes}
\label{sec:schemes}
\begin{figure}[t]
	\centering
	\includegraphics[width=.8\columnwidth]{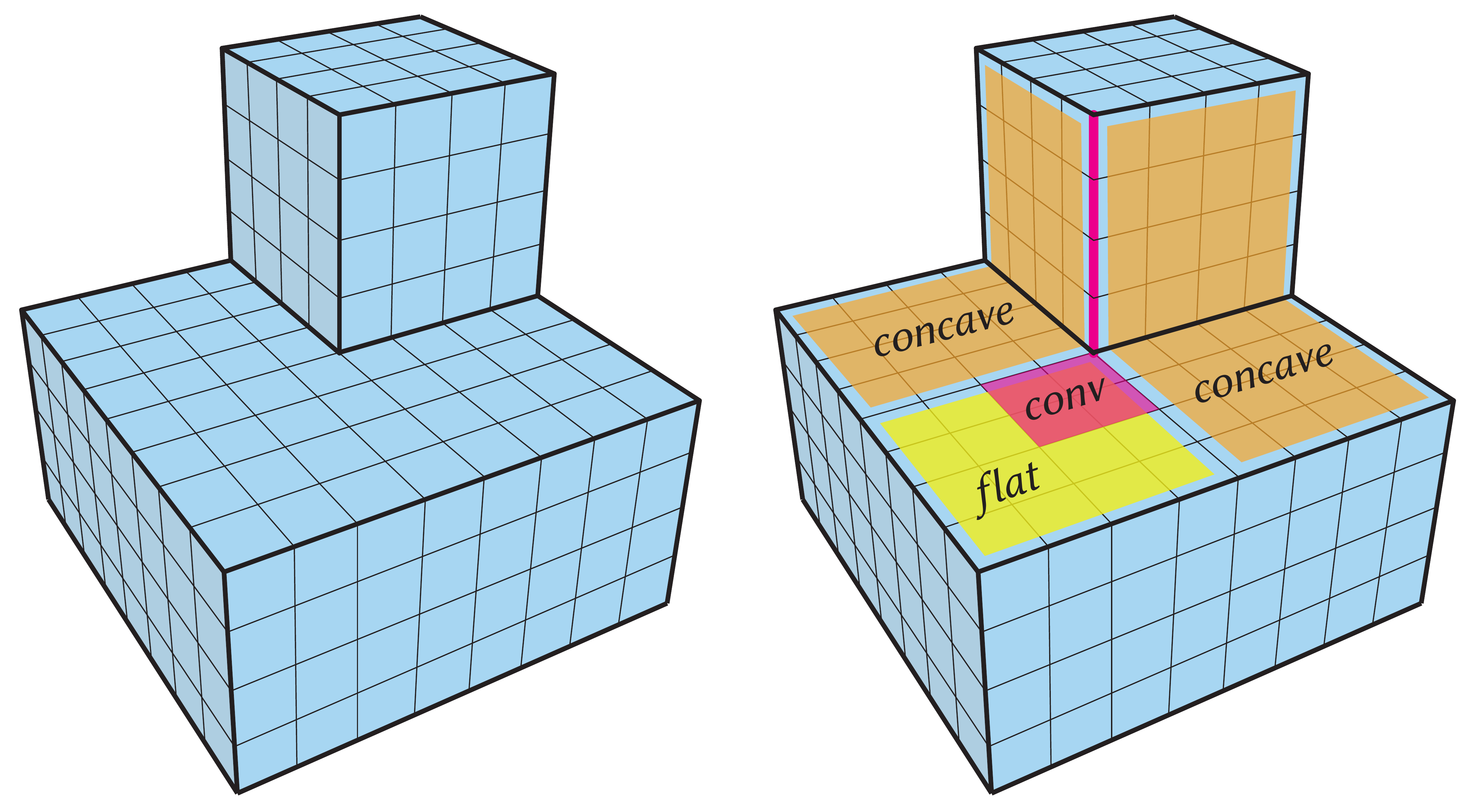}
	\caption{A non trivial transition involving a flat area, two concave edges, and one convex edge. The basis of the flat and convex schemes conflict (right), therefore basic transitions cannot be used directly, but must be combined in order to produce hybrid schemes that adapt to the local shape of the grid.}
	\label{fig:hybrid_trans}
\end{figure}

The basic transitions shown in the previous section cannot be directly used to transform an adaptively refined grid into a pure hexahedral mesh. As shown in Figure~\ref{fig:hybrid_trans}, many local configurations will necessitate hybrid transitions, which are a blend of the atomic patterns designed for the flat, convex and concave cases.

\begin{figure*}[t]
	\centering
	\includegraphics[width=\linewidth]{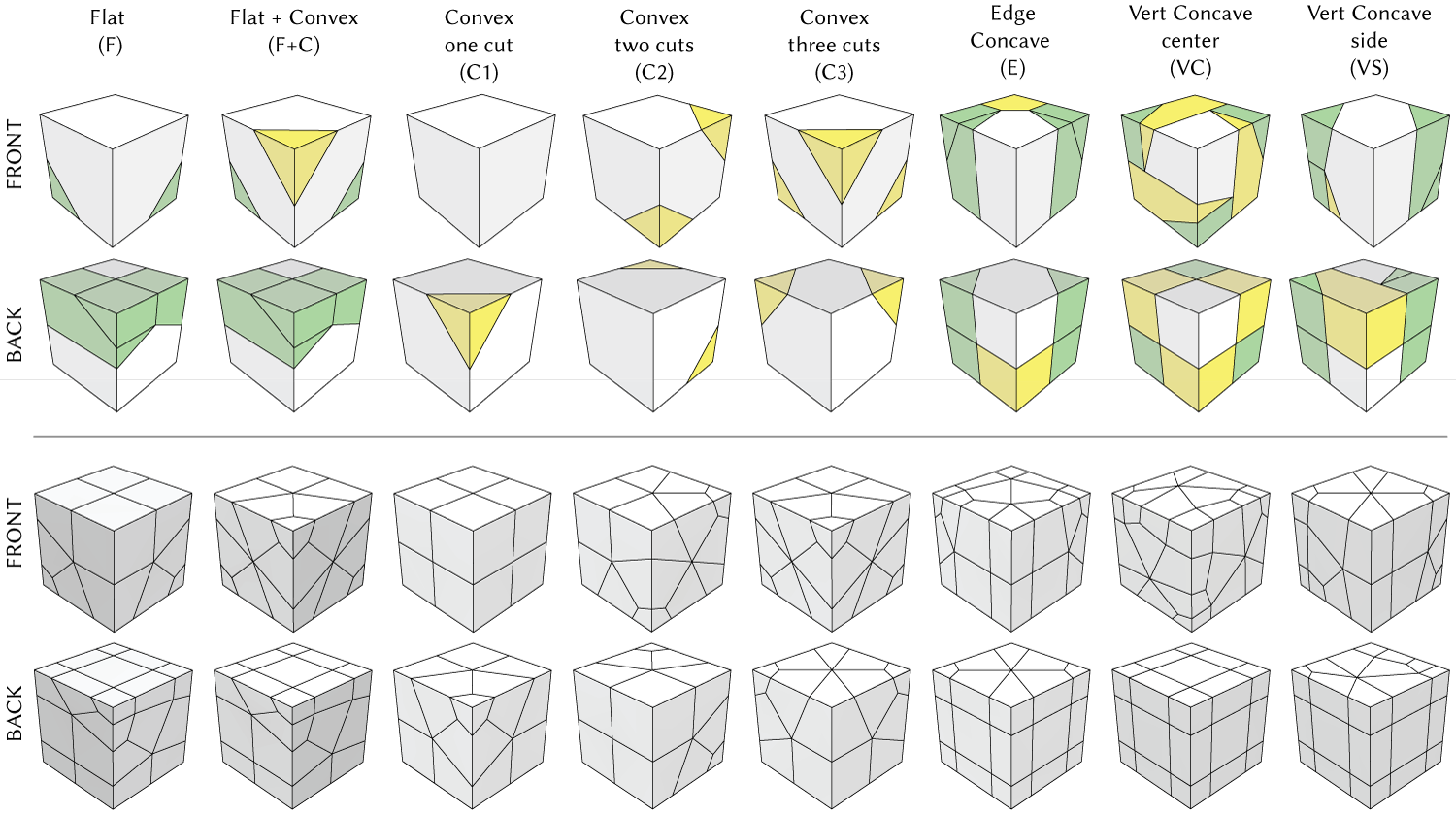}
	\caption{The 8 atomic schemes used to mesh all the transitions listed in Figure~\ref{fig:20cases}. Top: elements are color-coded with respect to their type. Green elements belong to prismatic chains that suppress the hanging nodes of a refined cluster. Yellow elements allow the green chains to bend around convex and concave edges. White elements are lids that fill the remaining volume. Bottom: hexahedralized transitions obtained with standard mesh dualization.}
	\label{fig:8schemes}
\end{figure*}

\begin{figure*}[t]
	\centering
	\includegraphics[width=\linewidth]{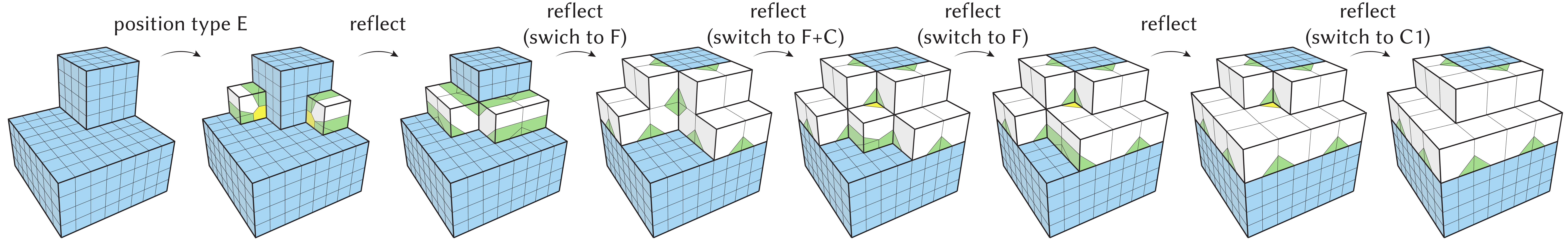}
	\caption{Example of installation of the atomic schemes in Figure~\ref{fig:8schemes} for a complex case involving flat, convex, and concave regions. All the necessary transitions can be realized with a combination of rigid movements and reflections of the basic schemes. The installing sequence is not mandatory, and any alternative sequence would provide the same result.}
	\label{fig:install}
\end{figure*}

\begin{figure*}[t]
	\centering
	\includegraphics[width=.9\linewidth]{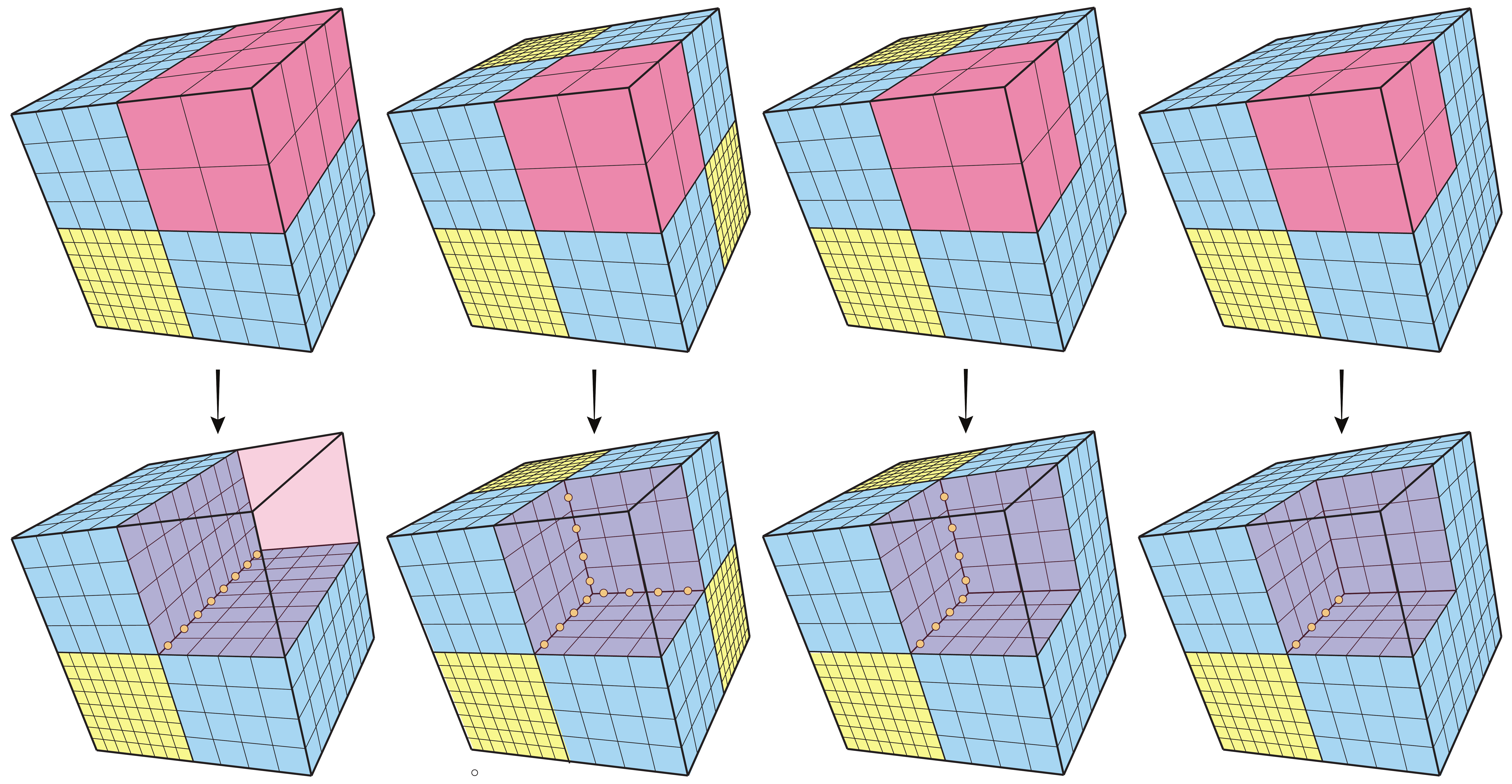}
	\caption{Weakly balanced grids may exhibit configurations where cells with three different levels of refinement are incident to the same edges, generating new hanging vertices that cannot be suppressed with prior schemes (yellow dots). Each column shows one of the four possible cases: elements can all be incident to the same concave open line, or to all (or a subset) of the three concave lines that terminate in a concave corner. As can be noticed hanging vertices belong to the finest sub-grids, and their suppression demands a blend between a convex transition (for the yellow part) and a concave transition (for the blue part).} 
	\label{fig:mul_ref}
\end{figure*}

To enumerate all the possible local configurations that may arise, we can imagine working in a grid where only two alternative refinement levels are possible: coarse and fine. Note that this hypothesis is not restricting: all grid-based methods assume that the adaptive grid is \emph{balanced}, which means that any couple of face, edge, or vertex adjacent cells can differ by at most one level of refinement~\cite{marechal2009advances}. It follows that indeed -- at a local level -- only two levels of refinement are possible.

Let us imagine having a cube split into 8 octants, and filling each octant either with a coarse $2 \!\times 2 \!\times 2$ or with a finer $4 \!\times 4 \!\times 4$ sub-grid. Since for each octant there are two possible choices, there exist $2^8$ alternative assignments. Ignoring symmetries and removing the two fully regular grids obtained by filling all octants with the same element, we obtain a set of 20 unique configurations, which correspond to all the possible transitions that may arise in a balanced adaptive grid (Figure~\ref{fig:20cases}). This combinatorial space is equivalent to the one explored by primal approaches (Section~\ref{sec:related}), and is also equivalent to that of Marching Cubes~\cite{lorensen1987marching}, which associates a sign to the cube's corners and obtains the same 20 possible binary assignments. This relation is even clearer in the dual version of MC~\cite{nielson2004dual}, which shows a lookup table that, up to a volumetric interpretation,  is equivalent to ours. Similar surface schemes had already been introduced in the Cuberille algorithm~\cite{herman1979three,chen1985surface}.

Implementing a transition scheme for each of the 20 patterns in Figure~\ref{fig:20cases} is, therefore, the simplest way to have a lookup table that exhaustively addresses all the cases. Note that this number is much bigger than the three schemes often presented as exhaustive in previous literature, and the reason is that prior works did not present the actual schemes, but just a particular instance of the basic transitions in Section~\ref{sec:basic_scheme}. 

To reduce the amount of configurations to the minimum, we present here an alternative method we used to encode the patterns. The basic idea is that many of the 20 schemes share some component, sometimes exactly as it is, some other times up to a rotational and reflection degree of freedom. We exploit this redundancy to define a minimum set of 8 atomic elements, which never overlap, and can be grouped together to reproduce all the 20 possible transitions. The full set is depicted in Figure~\ref{fig:8schemes} and comprises: one flat element (F); one hybrid flat and convex element (F+C); three convex elements (C1, C2, C3) that handle 1, 2 or 3 prismatic chains incident to the same grid cell; one element for concave edges (E), and two elements for concave vertices, once for the center (VC) and one for its sides (VS). The basic transitions discussed in Section~\ref{sec:basic_scheme} can be reproduced by considering simple arrangements of these 8 elements. As an example, the flat scheme in Figure~\ref{fig:flat} is composed of four elements of type F which can be positioned by starting from one of them and reflecting it four times across one of its lateral faces. Similarly, all the transitions shown in Figure~\ref{fig:20cases} can be obtained by compositions of the same 8 atomic elements. 

A pictorial illustration of the installation process is shown in Figure~\ref{fig:install}. Note that the sequence of operations is not mandatory. Since these atomic blocks do not conflict with each other, one can start by positioning a single brick, and simply proceed by placing the subsequent ones so as to preserve mesh conformity, always obtaining the same result. 




\section{Weakly balanced grids}
\label{sec:multiple_refinements}
All known methods for grid-based adaptive hexmeshing require that the input grid is \emph{strongly} balanced, which means that cells that differ by more than one level of refinement must not share any vertex, edge, or face. In this section we discuss a minimal extension of our basic schemes, which allows relaxing this stringent requirement, embracing a broader class of input grids and ultimately permitting us to obtain much coarser hexahedral meshes for same geometric accuracy. 

\begin{figure}
	\centering
	\includegraphics[width=\columnwidth]{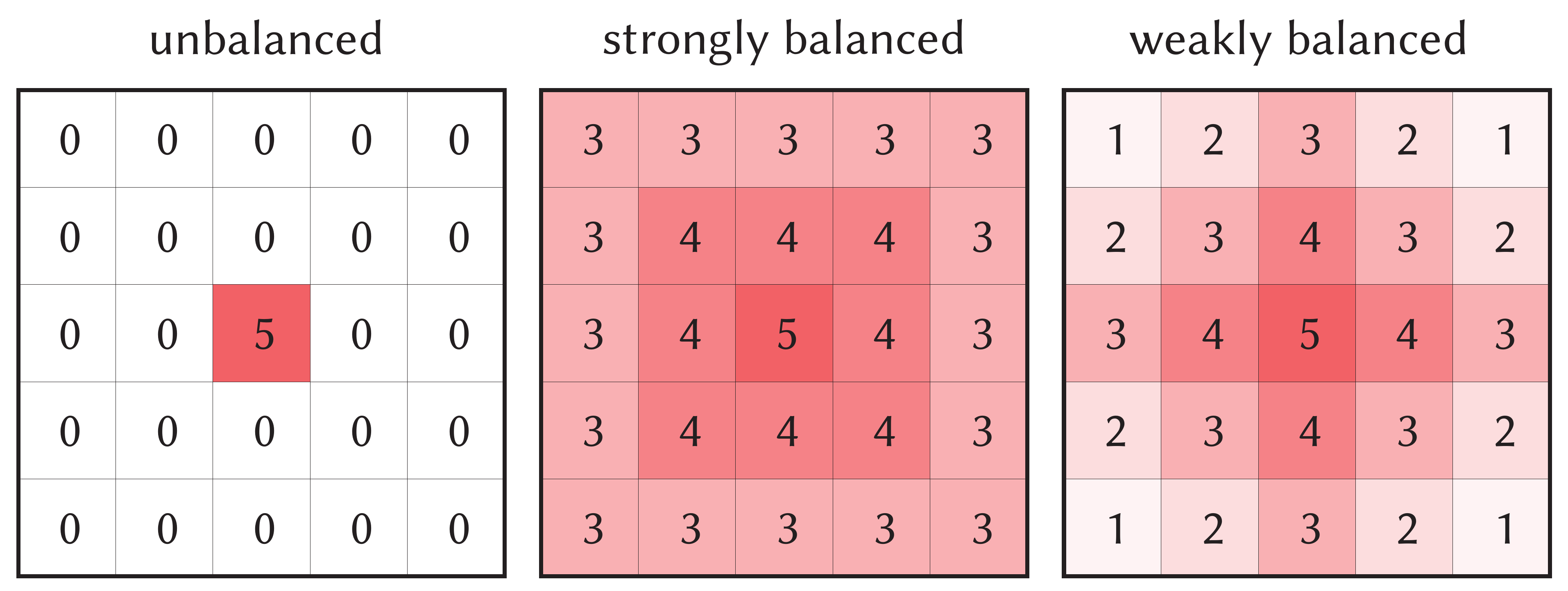}
	\caption{Starting from an unbalanced grid (left), fulfilling \emph{strong balancing} demands 80 steps of extra refinement (middle). If \emph{weak balancing} is permitted, the amount of necessary refinement is reduced by 25\%.}
	\label{fig:balancing}
\end{figure}

Our key observation is that when there is high disparity in the refinement associated to nearby cells, satisfying the strong balancing criterion requires a conspicuous amount of additional refinement, significantly increasing the cell count. Conversely, if the balancing criterion was \emph{weaker}, meaning that restrictions applied only to face-adjacent cells, the amount of necessary subdivisions would be much lower (Figure~\ref{fig:balancing}). 

\begin{figure}[h]
	\centering
	\includegraphics[width=\columnwidth]{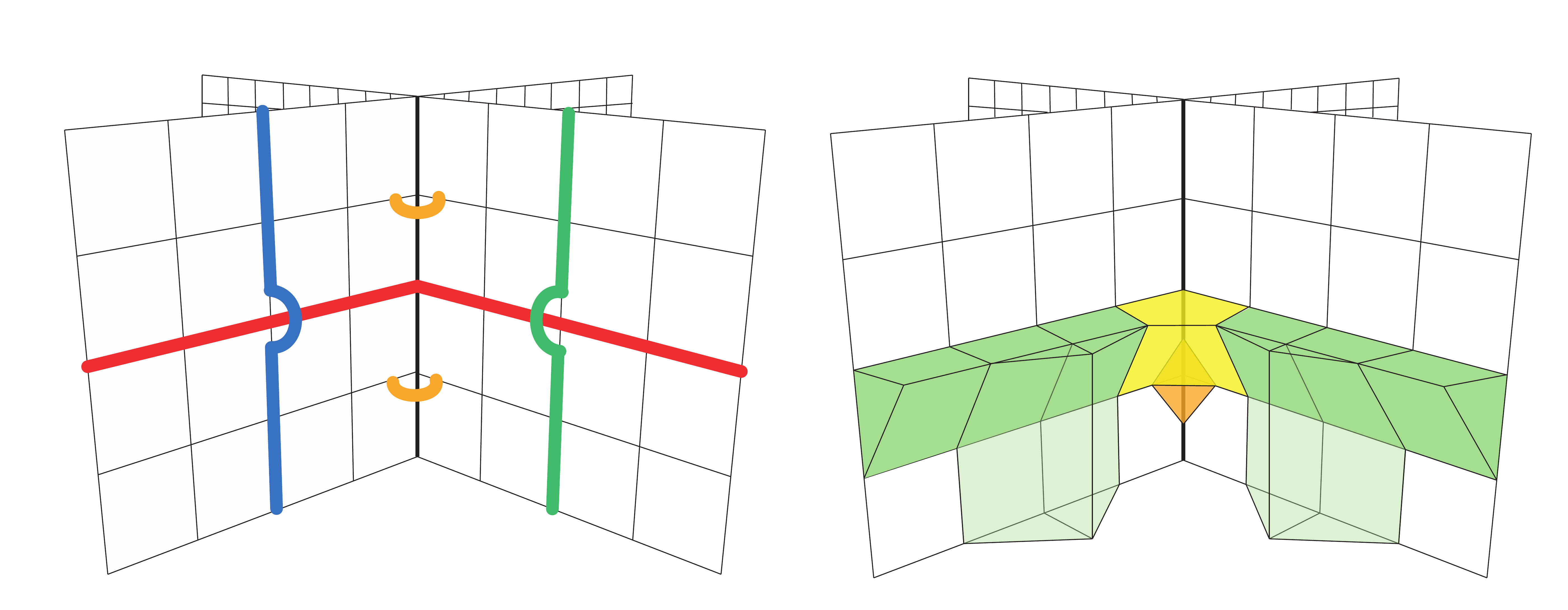}
	\caption{Left: hybrid convex/concave transition involving three different levels of refinement. Besides the three canonical chains of prisms of a standard concavity (in red, blue, green), there are two extra chains that take a convex turn around the concave edge (orange). Right: the two tetrahedral elements that ensure the convex transition (orange) partially overlap with the concave transition (yellow). As a result, the yellow faces -- that were pentagons in the basic concave transition -- become hexagons.}
	\label{fig:mul_ref_trans}
\end{figure}

\begin{figure}[h]
	\centering  
	\includegraphics[width=\columnwidth]{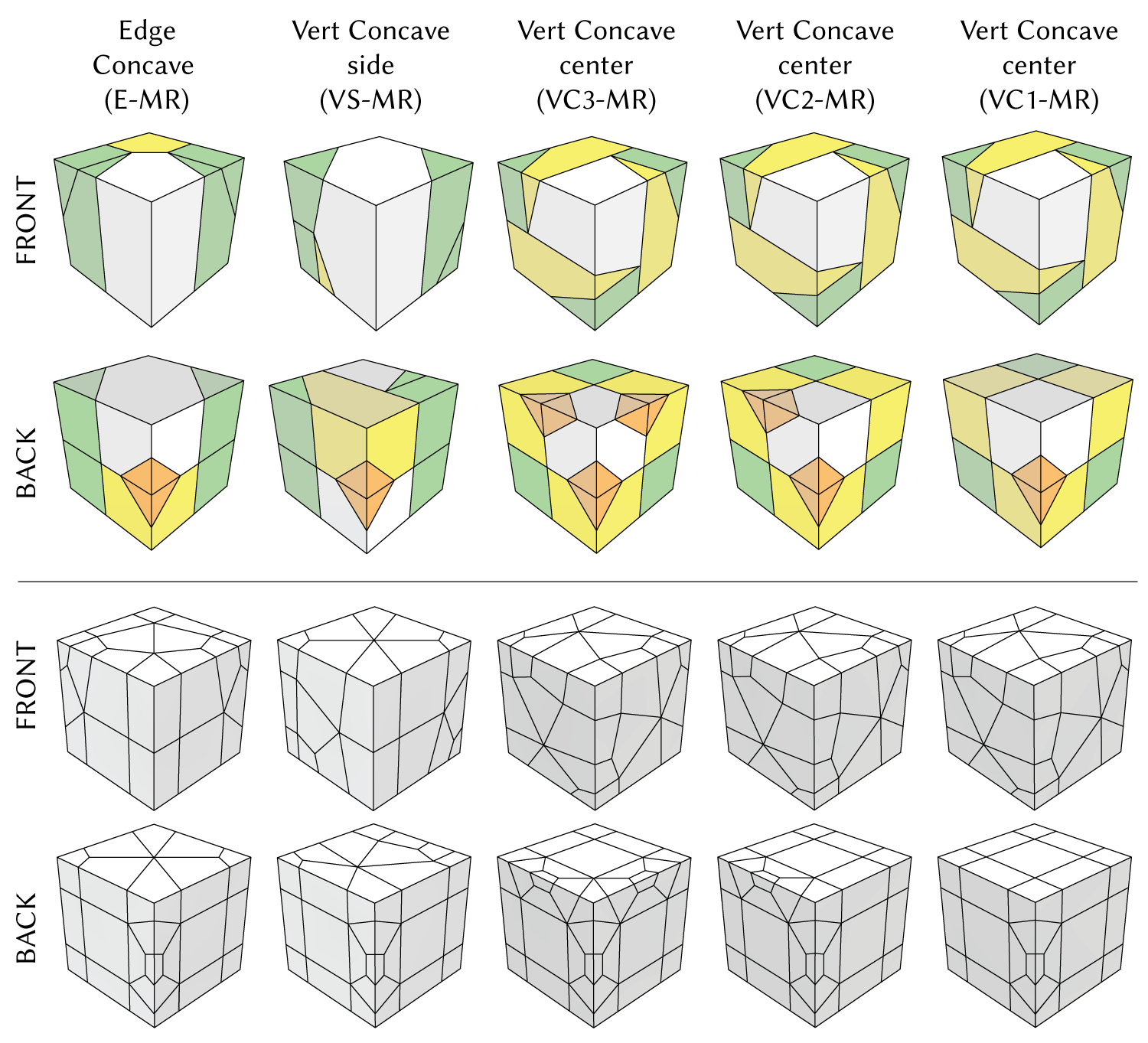}
	\caption{Hybrid concave/convex schemes to handle weakly balanced grid having cells with three levels of refinement incident to the same grid edges. Top: green elements are pieces of the prismatic chains. Yellow elements allow concave bending. Orange elements allow convex bending. White cells are lids to complete the volume. Bottom: hexahedralized transition blocks obtained with standard mesh dualization.}
	\label{fig:5extra}
\end{figure}

Weakly balanced grids may contain edges shared between cells with three different levels of refinement, and vertices incident to cells spanning four different levels of refinement. Luckily, the vertex case does not require any special handling because -- regardless of the size disparity -- any grid vertex has 8 incident cells and 6 incident edges, which means that it always yields a hexahedron in the dual mesh. This is also the reason why weakly balanced grids in 2D do not necessitate dedicated schemes. 
%
Conversely, edges shared by cells spanning three levels of refinement generate hanging vertices that must be incorporated into the mesh connectivity. As shown in Figure~\ref{fig:mul_ref} there are four possible cases, which correspond to an open concave edge, or a concave corner where 1, 2 or 3 of the incident concave edges contain additional hanging vertices. 

It is interesting to notice that the concave edges where the additional hanging nodes arise are convex edges for the (twice more refined) grid minors opposite to the concavity. This means that the schemes we need are essentially a blend between the basic convex and concave schemes shown in Figure~\ref{fig:8schemes}. A pictorial illustration of how to realize this blend for open concave edges is shown in Figure~\ref{fig:mul_ref_trans}. Note that the tetrahedra that realize the convex transition are located across the polygonal faces that permit the concave bending, transforming them from $n-$gons to $(n+1)-$gons. Specifically, the transition for concave edges required the use of pentagonal faces, which now become hexagons. The transition for concave corners required the use of hexagons, which now become heptagons. 
In terms of output results, this means that 
weakly balanced grids can be transformed into pure hexahedral meshes with singular edges of valence 3, 5, 6, and 7 using the 8 schemes in Figure~\ref{fig:8schemes}, plus 5 additional schemes shown in Figure~\ref{fig:5extra}.

\begin{figure*}
	\centering
	\includegraphics[width=\linewidth]{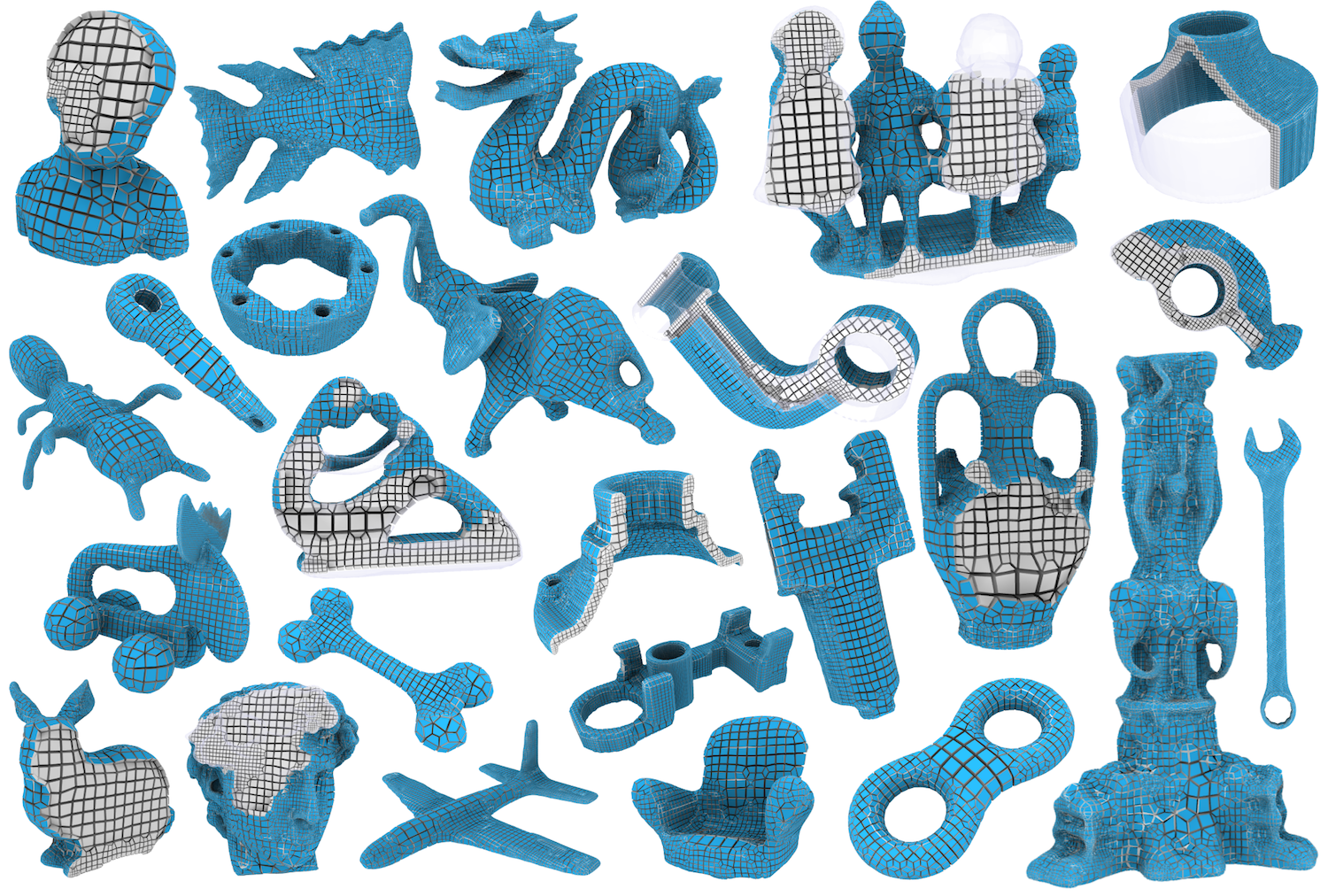}
	\caption{A random subset of the 202 models used to validate our method.}
	\label{fig:mosaic}
\end{figure*}

\section{Discussion}
\label{sec:discussion}
We implemented the whole scheme set and the software necessary to process an input grid in C++, releasing the code within the MIT licensed library CinoLib~\cite{Liv19}. Specifically, the 8+5 atomic elements whose combination realizes all the possible schemes for strongly and weakly balanced grids are hardcoded as general polyhedral meshes. The installation code is designed to translate, rotate, reflect and scale the atomic blocks, reproducing any desired transition. The so generated polyhedral meshes can be readily transformed into pure hexahedral meshes with standard mesh dualization, which is also available in the same library.

To validate our approach we hexmeshed the dataset released with~\cite{gao2019feature}, which comprises 202 organic and CAD models (Figure~\ref{fig:mosaic}). For each model we initialized an octree having a single node containing the bounding box, and then iteratively split octants intersected by the input surface until the grid size was at least twice as big as the local thickness of the shape, measured with the SDF~\cite{shapira2008consistent}. Similarly to previous methods, we limited the depth of the tree to 7 levels to avoid overly big meshes. We obtained a set of adaptive grids with hanging nodes which -- in the general case -- are incompatible with hexahedral meshing. We applied additional refinement to satisfy the topological criteria that are necessary to apply the transition schemes, producing two new grids for each shape, one strongly balanced and one weakly balanced. We obtained the final hexmeshes by: (i) applying the transitions schemes; (ii) dualizing the grid; (iii) removing the hexahedra not contained in the input shape; (iv) projecting the boundary vertices onto the target geometry. In the case of projection artifacts we post processed the output mesh with an off the shelf untangler~\cite{livesu2015practical}. Figure~\ref{fig:teaser} shows the main steps of the pipeline. 
Note that the focus of this work is on the transition schemes, and this is just a simplistic workflow. Scientific literature offers various alternative choices for octree splitting rules, feature preservation, and robust surface projection. Our approach can be combined with any of the existing techniques to obtain a fully-fledged meshing pipeline.\\

Regarding the impact of our schemes in the singular structure of the hexahedral mesh, in Table~\ref{tab:topology} we detail the valence of the primal faces for each of the proposed schemes. Strongly balanced grids only require singular edges with valence 3, 5, and 6. Weakly balanced grids may also require valence 7 singular edges, which appear when 4 out of the 5 additional schemes are used. Since our schemes can be used as is, without further modification, no edges with valence different from the ones declared here are possible.



\begin{figure}
	\centering
	\includegraphics[width=\columnwidth]{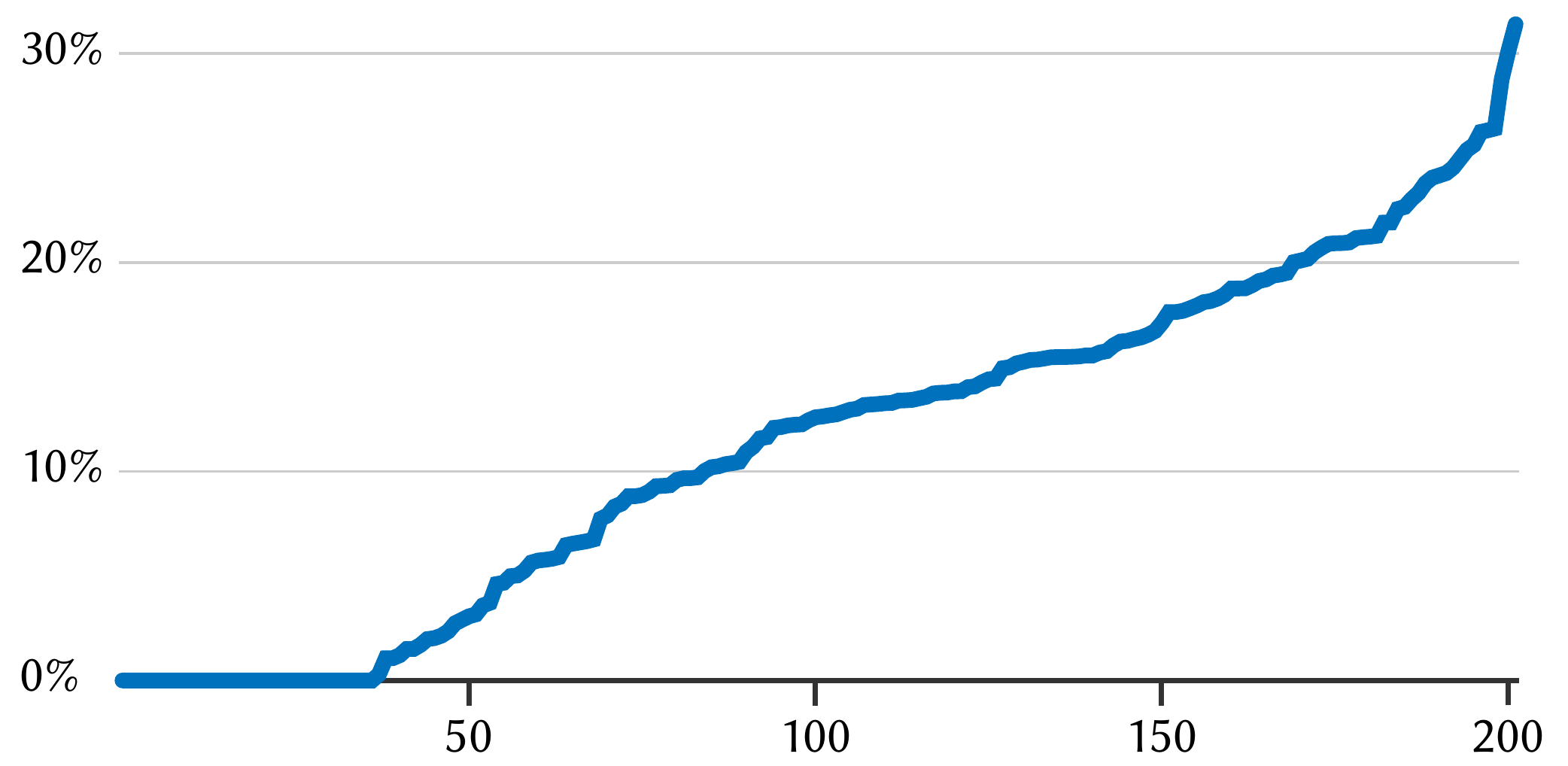}
	\caption{Considering the 202 hexahedral meshes in the dataset released with~\cite{gao2019feature}, refining an adaptive grid to fulfill only the weak balancing criterion allows to reduce the number of elements in more than 75\% of the cases. The average gain is around 15\% less grid elements, with peaks that go beyond 30\%. }
	\label{fig:gain}
\end{figure}

\subsection{Strong vs weak balancing}
\label{sec:bal_comparisons}
As mentioned before, in the general case an adaptive grid that has been split to faithfully approximate a target geometry does not fulfill the topological criteria that permit the application of the hexing schemes. Therefore, a certain amount of additional refinement must be applied to secure hexmeshability. In Figure~\ref{fig:gain} we compare cell count between pairs of adaptive grids that have been refined to fulfill strong and weak balancing for the 202 meshes in our testing dataset. As can be noticed, the 5 additional schemes introduced in Section~\ref{sec:multiple_refinements} permit to reduce the cell count in more than 75\% of the cases, while still being able to create a pure hexahedral mesh. Denoting with $\vert H \vert_S$ and $\vert H \vert_W$ the amount of cells in a strongly and weakly balanced grid, respectively, we measured mesh growth as $1 - \vert H \vert_W / \vert H \vert_S$. On average, we observed that weakly balanced grids allow to save around 15\% of the total amount of cells that a strongly balanced grid would require, and in some cases the gain grows up to more than 30\%. We can conclude that using weak balancing is highly convenient, as it allows to significantly reduce the number of hexahedra without sacrificing accuracy, because the amount of saved refinement was there only for topological reasons. 

\subsection{Comparisons with~\cite{marechal2009advances,gao2019feature}}
\label{sec:comparisons}
The most natural comparison for our set of schemes is with the original approach proposed by Marechal in~\shortcite{marechal2009advances} and with the alternative set of schemes recently proposed in~\cite{gao2019feature}. We attempted to implement both papers, but due to missing information in the articles we were never able to obtain a full prototype that could handle all the 20 transitions shown in Figure~\ref{fig:20cases}.
In his article, Marechal describes how to construct the flat scheme and the basic crossing of the prismatic chains, but he did not provide details on how this can be modified to enable bending and winding around concave corners. Similarly, Gao and colleagues describe the basic building block that allows handling the flat crossing (\emph{frustum} in their terminology) and also provide three additional schemes to account for adjacent frustums that form a flat, convex, and concave open angle. Concave corners are not taken into account, as well as conflicts that arise between basic schemes (see e.g., Figure~\ref{fig:hybrid_trans}). Restricting to the schemes shown in their paper, only 7 out of 20 possible cases shown in Figure~\ref{fig:20cases} can be handled. Contacting the authors, we could verify that indeed their software can handle all 20 possible cases, but this requires a non-trivial combination of the original schemes that is not mentioned in the article. We were not able to reproduce the results by ourselves, and despite the source code being publicly available, we still fail to understand how the mechanism works. 

\begin{table}[]
	\centering
	\resizebox{\columnwidth}{!}{%
		\begin{tabular}{|l|c|c|c|c|c|c|}
			\hline
			Type & ID & $F_3$ & $F_4$ & $F_5$ & $F_6$ & $F_7$ \\ \hline
			Flat 									& F 		  & 5 & 10 & 5 & - &  -\\
			Flat + Convex 						& F+C    & 9 & 9 & 4 & 2 &  -\\
			Convex 1 cut 				& C1 	  & 4 & 3 & 3 &  - &  -\\
			Convex 2 cuts 				 & C2 	 & 8 & 1 & 4 & 1 &  -\\
			Convex 3 cuts 			 	 & C3	 & 12 & - & 3 & 3 &  -\\
			Concave edge			  	 & E 	 & 8 & 14 & 7  &  1 &  -\\
			Concave vertex central	  & VC 	& 9 & 21 & - & 6 &  -\\
			Concave vertex side			& VS & 8 & 16 & 3 & 3  &  -\\ 
			\hline
			Concave edge 			& E-MR & 14 & 10 & 10 & 2 &  -\\
			Concave vertex central & VC1-MR & 16  & 17  & 4 & 5 & 1 \\
			Concave vertex central & VC2-MR & 23 & 14 & 6 & 5 & 2 \\
			Concave vertex central & VC3-MR & 30 & 12 & 6 &  6 & 3   \\ 
			Concave vertex side    & VS-MR & 15 & 12 & 7 & 2  & 1  \\
			\hline
		\end{tabular}%
	}
	\caption{Topological details for all the transition schemes proposed in the article. For each scheme we report the number of polygonal faces they contain. Once dualized, each polygon translates to an edge in the hexmesh with valence corresponding to the number of sides in the primal (the subscript $i$ in the $F_i$ notation).}
	\label{tab:topology}
\end{table}


Considering all these difficulties, we based our comparison on the data released with the publication of~\cite{gao2019feature}, which comprises 202 organic and CAD models hexmeshed both with the software released with~\cite{gao2019feature} and with the commercial implementation of~\cite{marechal2009advances}. Since these methods do not support weakly balanced grids and the software use different geometric criteria for refinement, results cannot be compared in terms of density or number of elements, but only in terms of singular structure.
We studied the valence of the singular edges produced by the three methods, which directly impacts mesh quality (Figure~\ref{fig:valence}). As expected, all meshes produced with our method contained edges with valence in between 1 and 6 (edges with valence lower than 3 are on the surface). For~\cite{marechal2009advances} edge valences were in the range $[2,9]$, and for~\cite{gao2019feature} were in the range $[2,10]$. Note that both software apply one padding layer to the outer surface, thus increasing valence for all edges having at least one vertex on the surface, or both vertices one step beneath the surfae.
Excluding from our analysis all such edges, the range becomes $[3,8]$ for both methods, still worse than what our schemes produce. 
Our analysis also revealed some interesting differences between these two methods. In fact, the schemes proposed in~\cite{gao2019feature} make heavy use of valence 7 edges, which appeared in 181 models out of 202, whereas the schemes proposed in~\cite{marechal2009advances} used valence 7 edges only in 23 models. This suggests that Marechal's schemes are very close to ours, although for some configurations they still require singular edges with valence above 6. Since the authors did not make their schemes explicit, it is impossible to say under what circumstances all these configurations arise.


\section{Conclusions}
\label{sec:conclusions}
We have extensively studied the topological schemes that permit to transform an adaptively refined grid into a pure hexahedral mesh. Previous literature had already proved that directly incorporating hanging nodes into the hexahedral mesh is not always possible. Therefore, our analysis restricted to dual schemes, which aim to generate a general polyhedral mesh that yields only hexahedral elements when dualized.

We started our study from the seminal work of Marechal~\shortcite{marechal2009advances}, who pioneered dual approaches. We have shown that both his schemes and the schemes proposed in later articles are not exhaustive and do not contemplate ambiguities that arise when one tries to implement them. 
More precisely, we have shown that the combinatorial space of all possible configurations for strongly balanced adaptive grids is equivalent to that of Marching Cubes, and amounts to 20 different transitions. We also proposed a reduced set of 8 atomic transitions that -- combined with a set of rigid movements and reflections -- allow to span the whole combinatorial space while requiring less effort to the developer. 

From our point of view, the scheme set we derived is a strict interpretation of the original ideas of Marechal. Surprisingly enough, our output meshes seem to be different from the ones produced with his approach. We could not attempt a side by side comparison because Marechal did not reveal the full set of schemes, but the commercial implementation of~\cite{marechal2009advances} produces meshes that contain singular edges with valence up to 8, whereas we have shown that all possible transitions can be realized bounding edge valence to 6. To this end, our schemes also outperform the schemes recently proposed in~\cite{gao2019feature}, which reach maximum valence 8, and also seem to make much heavier use of valence 7 edges, which appeared in the vast majority of their results.


All in all, we can conclude that our schemes advance the state of the art, putting a tighter bound on the valence of irregular edges, and also permitting -- for the first time -- to process weakly balanced grids, a class of inputs that is not supported by the prior art. 

For future works, we believe that the major improvements are unlikely to come from a new set of better schemes for the same class of grids, but rather on novel ideas to embrace a broader class of input adaptive grids. Based on our understanding, the current bottleneck in the pipeline is the amount of refinement that adaptive grids must undergo to ensure the applicability of the hexing schemes.
Our extension to weakly balanced grids is a first step in this direction, and already proved its effectiveness on a vast set of shapes. We believe that more can be done in this regard, and we will devote our future efforts to working in this direction.

%
%
%
%


\section*{Acknowledgements}
We are deeply grateful to Xifeng Gao for the clarifications on his work, and for kindly hexmeshing for us the full set of transition schemes for strongly balanced grids. We also thank the authors of Hexalab~\cite{bracci2019hexalab}, which we extensively used for all the measurements and renderings in the article.

\bibliographystyle{ACM-Reference-Format}
\bibliography{biblio}

\appendix

\end{document}